\documentclass[aps,twocolumn,prb,showpacs,superscriptaddress]{revtex4}
\usepackage[dvips]{graphicx}
\usepackage{amssymb}
\usepackage{epsfig}
\usepackage{epsf}

\def\frac#1#2{{\textstyle{#1 \over #2}}}

\def\be{\begin{equation}} \def\ee{\end{equation}}
\def\bea{\begin{eqnarray}} \def\eea{\end{eqnarray}}

\def\nn{\nonumber}

\def\vk{{\vec k}}
\def\vkp{{\vec k}'}
\def\phik{\phi_{\vec k}}
\def\thek{\theta_{\vec k}}
\def\phikp{\phi_{{\vec k}'}}
\def\thekp{\theta_{{\vec k}'}}
\def\vq{{\vec q}}
\def\vp{{\vec p}}
\def\vr{{\vec r}}
\def\vS{{\vec S}}

\def\psid{\psi^\dagger}
\def\vpha{{\vphantom{\sum_N^N}}}

\begin{document}

\title{Quasiparticle Interference on the Surface of the
Topological Insulator Bi$_2$Te$_3$}

\author{Wei-Cheng Lee}
\affiliation{Department of Physics, University of California, San Diego, CA 92093}

\author{Congjun Wu}
\affiliation{Department of Physics, University of California, San Diego, CA 92093}

\author{Daniel P. Arovas}
\affiliation{Department of Physics, University of California, San Diego, CA 92093}

\author{Shou-Cheng Zhang}
\affiliation{Department of Physics, McCullough Building, Stanford
University, CA 94305}

\date{\today}

\begin{abstract}
The quasiparticle interference of the spectroscopic imaging scanning
tunneling microscopy has been investigated for the surface states of
the large gap topological insulator Bi$_2$Te$_3$ through the
$T$-matrix formalism.  Both the scalar potential scattering and the
spin-orbit scattering on the warped hexagonal isoenergy contour are
considered.  While backscatterings are forbidden by time-reversal
symmetry, other scatterings are allowed and exhibit strong
dependence on the spin configurations of the eigenfunctions at
$\vk$ points over the isoenergy contour.  The characteristic
scattering wavevectors found in our analysis agree well with recent experiment results.
\end{abstract}
\maketitle

\section{Introduction}
The theoretical proposal \cite{bernevig2006a,kane2005,bernevig2006,
qi2008,fu2007a,moore2007,roy2006,zhang2009a} and experimental
discovery of the topological insulators
\cite{koenig2008,hsieh2008,xia2009,chen2009} have provoked an
intensive research effort in condensed matter physics. Topological
insulators (TI) with time-reversal symmetry are generally
characterized by a topological term in the electromagnetic action
with a quantized coefficient\cite{qi2008}. These states have been
theoretically predicted and experimentally observed in both two and
three dimensions, including the two-dimensional (2D) HgTe/HgCdTe
quantum wells \cite{bernevig2006a, koenig2008}, and bulk
three-dimensional materials Bi$_2$Te$_3$, Bi$_2$Se$_3$ and
Bi$_{1-x}$Sb$_x$
\cite{zhang2009a,chen2009,xia2009,fu2007a,hsieh2008,roushan2009}.
They exhibit robust gapless modes at boundaries, {\it e.g.\/} a 1D
helical edge mode for 2D TIs, and a 2D helical surface mode for 3D
TIs with odd numbers of Dirac cones. Due to time reversal symmetry,
backscattering is forbidden for the helical edge and surface states,
and an analysis of interaction effects for the 1D helical edge modes
shows they are stable against weak and intermediate strength
interactions \cite{wu2006,xu2006}. Bi$_2$Te$_3$ and Bi$_2$Se$_3$
have been predicted to have bulk band gaps exceeding room
temperature\cite{zhang2009a}, which makes them promising for future
applications.

Zhang {\it et al} predict that the surface states of Bi$_2$Te$_3$
consist of a single Dirac cone at the $\Gamma$ point, and that the
Dirac cone evolves into a hexagonal shape at higher
energy\cite{zhang2009a}. Furthermore, near the Dirac point, the spin
of the electron lies perpendicular to the momentum. Angle-resolved
photo-emission spectroscopy (ARPES) measurements performed on the
surface of  Bi$_2$Te$_3$ have confirmed these predictions in
detail\cite{chen2009,hsieh2009}. The typical shape of the Fermi
surface is a snowflake-like warped hexagon. The low-energy
\textsf{O}(2) symmetry of the Dirac cone is broken due to the
$C_{3v}$ symmetry of the underlying lattice\cite{zhang2009a}, and
can be modeled by a warping term in the effective
model\cite{fu2009}. Another powerful surface probe, spectroscopic
scanning tunneling microscopy (STM), is sensitive to quasi-particle
interference (QPI) around impurities, and provides an important tool
to study electronic structures in unconventional materials, such as
high T$_{\rm c}$ cuprates \cite{hanaguri2007,wang2003}. It can
provide information in momentum space through real space measurement
with a high energy resolution. Recently, several groups  have
performed STM measurements on surface states of Bi$_2$Te$_3$ and
Bi$_{1-x}$Sb$_x$
\cite{roushan2009,alpichshev2009,zhang2009,gomes2009}.
Backscattering induced by non-magnetic impurities between
time-reversal (TR) partners with opposite momenta is forbidden due
to their opposite spin configurations.  This is confirmed by the
real space Friedel oscillation pattern and by analysis of the QPI
characteristic scattering wavevector.

In this paper, we perform a detailed QPI analysis of the surface states of the topological insulator
Bi$_2$Te$_3$.  A general TR-invariant impurity potential including scalar and spin-orbit scattering
components is studied using the standard $T$-matrix formalism.
The scattering on the iso-energy surface strongly depends on the both momentum and
spin orientation.  Scattering between TR partners vanishes as a consequence of TR symmetry.
The scattering is dominated by wavevectors which connect regions on the Fermi surface of
extremal curvature, but also accounting for spin polarization.  
STM experiments\cite{alpichshev2009,zhang2009} have yielded rich information about the QPI structure. 
In addition to the absence of backscattering, the STM experiments 
also observed recovered scattering\cite{alpichshev2009} at a wavevector ($\vec{k}_{nest}$ in their, and $\vec{q}_2$ in our notation),
and an extinction\cite{zhang2009} (i.e. near absence
of scattering) ($\vec{q}_3$ in their and our notation), both at wavevectors which do not connect TR states.
Below, we offer a novel explanation
of this experimental puzzle.  Our results are in excellent overall agreement with the QPI experiment in Bi$_2$Te$_3$.

\section{Suface Dirac model with warping term}
\begin{figure}
\includegraphics{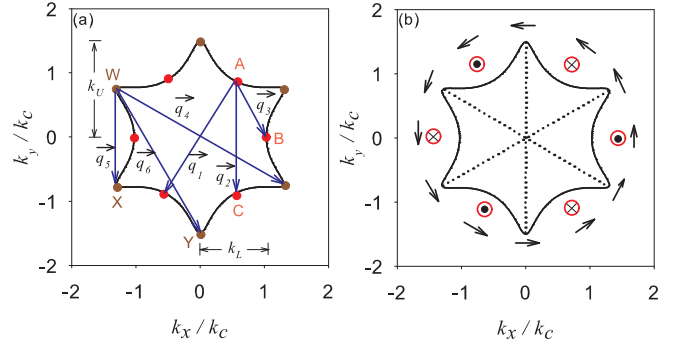}
\caption{\label{fig:fs} (Color online) (a)The iso-energy contour
near the $\Gamma$ point for $E=1.5 E^*$ with snow-flake shape.
The $\hat{x}$ and $\hat{y}$ axes are chosen to be the $\Gamma$-$K$
and $\Gamma$-$M$ directions respectively, and $k_c=\sqrt{v/\lambda}$.
The red and brown (dark gray) dots refer to the valley and the tip
points on the contour, and the arrows indicates six
representative scattering wavevectors. $k_L$ and $k_U$ are solutions
of $E_+(k_L,\theta=0)=E_+(k_U,\theta=\pi/2)=E$ which are the
boundary of the truncation for the $\vec{k}$-integration used
in this paper.
(b) The spin orientations of the eigenfunctions for $\alpha_+$ band at
valley and tip points. The dotted lines refer to the
mirror-symmetric lines ($\Gamma$-M), and the system has a
three-fold rotational symmetry.
The arrow indicate the spin configuration in the $xy$ plane and
the solid circle (cross) refers to $S_z$ being along $+\hat{z}$ ($-\hat{z}$).
At the cusp points the spin lies only on the $xy$ plane while
$S_z$ has the largest magnitude at the valley points
with staggered signs.}
\end{figure}

The $\vec{k}\cdot\vec{p}$ Hamiltonian for the surface Dirac cone was
first derived in Ref. \onlinecite{zhang2009a}.  The bare Hamiltonian is written as ${\cal H}_0=\int\!d^2\!k\,
\psi^\dagger(\vk)\,H(\vk)\,\psi(\vk)$, where
$\psi^\dagger(\vec{k})=(c^\dagger_{\vec{k}\uparrow},c^\dagger_{\vec{k}\downarrow})$.
With the addition of the cubic warping term \cite{fu2009}, 
\be H(\vk)=v\big(\vec{k}\times\vec{\sigma}\big)\cdot\hat{z} +\lambda
k^3\cos3\phik\>\sigma^z\ . \ee
The azimuthal angle of $\vk$ is $\phik=\tan^{-1}(k_y/k_x)$,
where the $\Gamma$-$K$ direction is taken as $\hat{x}$ axis.
Following Ref. \onlinecite{fu2009}, the
quadratic terms are dropped since they do not significantly change the 
shape of the constant energy contour, and the characteristic energy
and wavevector scales are defined as: $E^*=v\,k_c$ and
$k_c=\sqrt{v/\lambda}$. This Hamiltonian can be diagonalized by
introducing
\bea \hat{U}(\vk)=\left(
\begin{array}{cc}
\cos(\thek/2)  &  i e^{-i\phik}\sin(\thek/2) \\ & \\
i e^{i\phik}\sin(\thek/2) & \cos(\thek/2) \\
\end{array}\right)\ ,
\label{eigenfunctions}
\eea
where
$\tan \thek=k^2_c/( k^2\cos 3\phik)$.   One then finds $H(\vk)=E(\vk)\,U(\vk)\,\sigma^z\,U^\dagger(\vk)$,
with eigenvalues $E_\pm=\pm E(\vk)$ where
\be
E(\vk)=\sqrt{(vk)^2 + (\lambda k^3\cos 3\thek)^2}\ .
\ee
In fig. \ref{fig:fs}(a) we plot the isoenergy contour $E=1.5\, E^*$, which
qualitatively reproduces the snowflake Fermi surface observed in the
first-principles calculation and the ARPES
experiment \cite{zhang2009a,chen2009,fu2009}.  As for the scattering
process, we take
\bea {\cal H}_{\rm imp}=\!\!\int \! d^2k\,d^2k'\>V_{\vk-\vkp}\,
\psid(\vkp)\left[{\mathbb I} + i c \,\vk\times\vkp
\cdot\vec{\sigma}\right]\psi(\vec{k}). \label{hs} \eea
For a single short-ranged scatterer we may approximate 
$V_{\vk-\vkp}\approx V_0$. The second term
corresponds to the spin-orbit scattering with the coefficient $c$
describing its relative strength to the potential scattering.
It is convenient to project the potential onto the eigenbasis of ${\cal H}_0$, so
\bea \hat{V}_{\vk,\vkp} \equiv V_0\>\hat{U}^\dagger(\vkp)
\left[{\mathbb I}+ i c\,\vk\times\vkp
\cdot\vec{\sigma}\right]\hat{U}(\vk). \label{vkk} \eea
For simplicity, we first consider the $c=0$ case (pure scalar potential scattering), returning
later to the general spin-orbit case ($c\neq 0$).  Since the spectrum is particle-hole symmetric,
let us focus on a definite (positive) sign of the energy.  The QPI will then be dominated by scatterings inside
the positive energy band, whose effective scattering potential is:
\be
\hat{V}_{\vk,\vkp}^{(11)}=V_0\bigg[\cos\frac{\thek}{2}\cos\frac{\thekp}{2}
+\sin\frac{\thek}{2}\sin\frac{\thekp}{2}\,e^{i(\phik-\phikp)}\bigg]\ .
\label{v11}
\ee
This effect also appears in the QPI analysis of the orbital-band systems
where orbital hybridization brings strong momentum dependence
to the scattering process \cite{lee2009}.

\section{Effect of spin orientation on the QPI pattern}
The points of extremal curvature on the Fermi surface are divided into two groups, arising from the `valleys' ($k=k_L$, positive curvature)
and `tips' ($k=k_U$, negative curvature).   We define the complexified points $A=k_L\,e^{i\pi/3}$, $B=k_L$, $C=k_L e^{-i\pi/3}$,
$W=k_U e^{5\pi i/6}$, $X=k_U e^{-5\pi i/6}$, and $Y=k_U e^{-i\pi/2}$.  Then from eqn.  \ref{v11} we obtain
$\big|V^{(11)}_{AB}\big|^2=\frac{3V_0^2}{4}\sin^2\vartheta$, $\big|V^{(11)}_{AC}\big|^2=\frac{V_0^2}{4}+\frac{3V_0^2}{4}\cos^2\vartheta$,
and $V^{(11)}_{A{\bar A}}=0$, where ${\bar A}=-A$, corresponding to scattering through the vectors $\vq_3$, $\vq_2$, and $\vq_1$,
respectively, with $\tan\vartheta=(k_c/k_L)^2$.  We also find $\big|V^{(11)}_{WX}\big|^2=\frac{3V_0^2}{4}$,
$\big|V^{(11)}_{WY}\big|^2=\frac{V_0^2}{4}$, and $V^{(11)}_{W{\bar W}}=0$.  These processes are depicted in fig. \ref{fig:fs}(a).

While $V^{(11)}_{A{\bar A}}=V^{(11)}_{W{\bar W}}=0$ is a direct consequence of TR symmetry, the other processes through scattering
vectors $\vq_{2,3,5,6}$ are in general finite.  Their amplitude variation may be understood in terms of the spin orientation of the eigenfunctions
throughout the Brillouin zone, ${\vec S}(\vk)=(-\sin\thek\sin\phik \,,\, \sin\thek\cos\phik \,,\, \cos\thek)$, depicted in fig. \ref{fig:fs}(b).
Bi$_2$Te$_3$ has the symmetry of $C_{3v}$, {\it i.e.\/} three-fold
rotational symmetry plus the three reflection lines ($\Gamma$-$M$ plus two equivalent lines).  Therefore at the tips $S^z(\vk)$
must vanish since $\sigma^z$ is odd under the mirror operation.  $S^z(\vk)$ has the largest magnitude at the valleys, but with
staggered signs, as shown in the figure. Since scalar potential scattering does not flip electron spin, its matrix element
is largest when $\vS(\vk)\cdot\vS(\vkp)$ is large and positive, {\it i.e.\/} high spin overlap. 
This echoes the experimental finding of Pascual {\it et al.}\cite{pascual} that in the QPI pattern on Bi(110), 
only the scattering processes preserving the spin orientation are visible. 
One major difference, however, betwwen Bi(110) and Bi$_2$Te$_3$ is that the former has multiple Fermi surfaces and the scattering 
processes preserving spin orientations do exist at finite $\vec{q}$, while the later only has one Fermi surface and therefore 
no such scatterings could exist.
At the tips, the spin lies in-plane, with $\thek=\frac{\pi}{2}$,
independent of the scanning energy $E$.  It can be checked that $\vS(\vk+\vq_5)\cdot\vS(\vk) > \vS(\vk+\vq_6)\cdot\vS(\vk)$, hence
$\big|V^{(11)}_{WX}\big|^2 > \big|V^{(11)}_{WY}\big|^2$.  For scatterings between the valleys, $\vS(\vk)\cdot\vS(\vkp)$
depends crucially on $S^z(\vk)$ and $S^z(\vkp)$.  Accounting for the valley-to-valley oscillation in $\vS(\vk)$, we conclude that as
the scanning energy increases, $\big|V^{(11)}_{AC}\big|^2$ grows while $\big|V^{(11)}_{AB}\big|^2$ shrinks.
This simple argument gives a qualitative explanation for the absence of the $\vq_3$ scattering in the STM experiment \cite{zhang2009}.
For typical experimental parameters \cite{fu2009}, $E/E^*\approx 1.5$ and $k_L/k_c\approx 1$.  In this case we estimate
the scalar potential scattering gives that
$\big|V^{(11)}_{WX}\big|^2 : \big|V^{(11)}_{AC}\big|^2 : \big|V^{(11)}_{AB}\big|^2 : \big|V^{(11)}_{WY}\big|^2 \approx 6:5:3:2$.


\section{Numerical Results}
To specifically compute the QPI image, we employ a $T$-matrix
approach \cite{balatsky2006} for multiband systems \cite{lee2009}.
In the operator basis $\Psi(\vk)=U(\vk)\,\psi(\vk)$, the Green's function is
written in matrix form as
\bea
\hat{G}(\vk,\vkp,\omega)&=&\hat{G}_0(\vk,\omega)\,
\delta_{\vk,\vkp} + \hat{G}_0(\vk,\omega)\,
\hat{T}_{\vk,\vkp}(\omega) \, \hat{G}_0(\vkp,\omega)\nn\\
\label{gtg} \eea
where the $T$-matrix satisfies
\bea \hat{T}_{\vk,\vkp}(\omega)=\hat{V}_{\vk,\vkp} +\int\!\!d^2p\>
\hat{V}_{\vk,\vp} \, \hat{G}_0(\vp,\omega) \,
\hat{T}_{\vp,\vkp}(\omega)\ , \label{tmatrix} \eea
and
$\big[\hat{G}_{0,\sigma}(\vk,\omega)\big]_{ab}=
\big[\omega+i\delta-E_a(\vk)\big]^{-1}\delta_{a,b}$ are the bare Green's
functions.  In spectroscopic imaging STM \cite{balatsky2006}, the
conductance ($dI/dV$) measured by the STM is proportional to the
local density of states defined as
\be \rho(\vec{r},\omega)=\rho_\uparrow(\vec{r},\omega)+\rho_\downarrow(\vec{r},\omega)\ , \ee
where $\rho_\sigma(\vec{r},\omega)={\rm Im}
G_\sigma(\vec{r},\vec{r},\omega)$ is the local density of states for
spin $\sigma$. The QPI image in the Brillouin zone
$\rho(\vec{q},\omega)$ is then obtained by performing the Fourier
transformation of the conductance $dI/dV$.  As a result, we can
calculate $\rho(\vq,\omega)$ using the $T$-matrix formalism by:
\bea
\rho(\vq,\omega) &=& \int \!\!d^2r \> e^{i \vq\cdot\vr}\,\rho(\vr,\omega) \nn\\
&=&{1\over 2i}\int \!\! d^2k\>\textsf{Tr} \bigg[\hat{U}(\vk)\,\hat{G}(\vk,\vk+\vq,\omega)\,
\hat{U}^\dagger(\vk+\vq) \nn\\
&&\quad -\Big(\hat{U}(\vk)\,\hat{G}(\vk,\vk-\vq,\omega)\,
\hat{U}^\dagger(\vk-\vq)\Big)^* \bigg]
\label{rhoq}
\eea
where the trace is taken with respect to the matrix index.
Because physically STM measures the local density of states in the spin basis of $\hat{\psi}(\vk)$, while our $T$-matrix theory here is developed
in the eigenbasis of $\hat{\Psi}(\vk)$, the $\textsf{SU}(2)$ rotation matrices $\hat{U}(\vk)$ are introduced in the last line of eq. \ref{rhoq}
to transform back to the physical spin basis.  Because the first term in eq. \ref{gtg}, $\rho(\vq=0)$ contains the sum of the total
density of states without the impurity, which makes it much larger than $\rho(\vq\neq 0)$,
we only plot $|\rho(\vq\neq 0)|$ in order to reveal weaker structures of the QPI induced by the impurity scattering.

\begin{figure}
\includegraphics{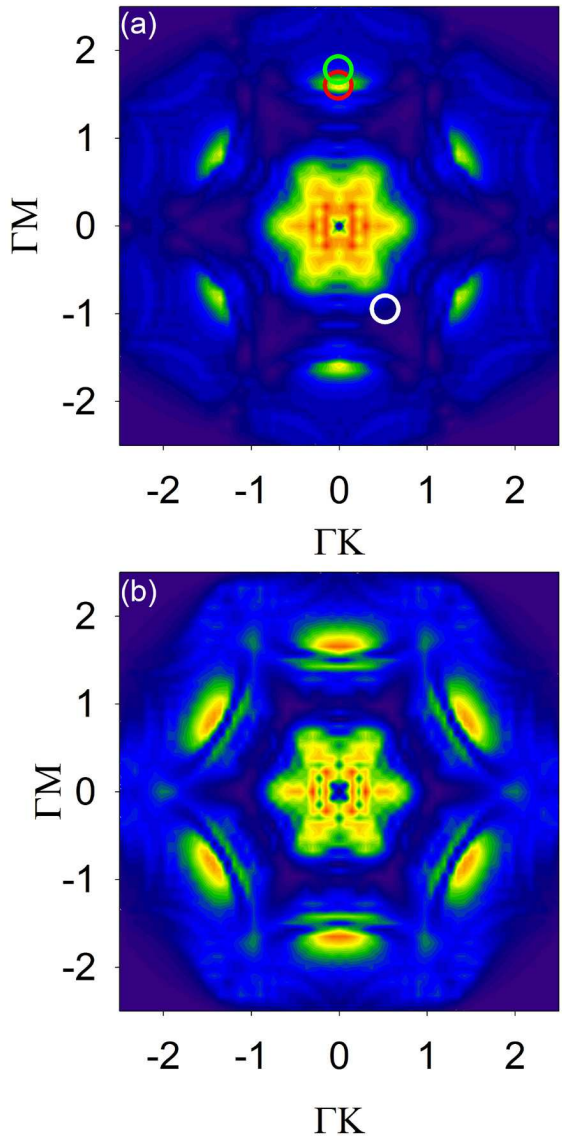}
\caption{\label{fig:qpi} (Color online) The quasiparticle interference image for
(a) $c=0$ and (b) $c=0.5$ with $E=1.5\, E^*$ and $V_0/E^*=0.1$. In
this case, $k_L/k_c=1.029$ and $k_U/k_c=1.5$.  (a) The strongest
large $\vq$ scatterings are $\vq_5$ and $\vq_2$ indicated by the red (dark gray) and green (light gray) circles
(and their symmetric points). $\vq_3$ (indicated by the white circle)
is too weak to be seen. (b) For $c=0.5$, new QPI features with large
momenta are visible.}
\end{figure}

We solve eq. \ref{tmatrix} numerically, using 2D polar coordinates.
Since the dominant scattering processes are between $\vk$ points on the constant energy contour $E_+(k,\theta)=E$
(we focus on $E>0$ here), we perform the integration within the range $k_L\leq k\leq k_U$ with $k_L$ and $k_U$ indicated in Fig.
\ref{fig:fs}(a). The resulting QPI images are plotted in fig. \ref{fig:qpi} for $c=0$ with $E=1.5\, E^*$ fixed.  For this choice of
parameters, $k_L/k_c= 1.029$ and $k_U/k_c=1.5$.  As shown in fig. \ref{fig:qpi}(a), $\vq_5$ and $\vq_2$
indicated by the red (dark gray) and green (light gray) circles are the strongest features while $\vq_3$ (indicated by the white circle) is almost invisible.
The reason why $\vq_5$ is even stronger than $\vq_2$ while they have comparable scalar scattering potential is due to the 
difference in the density of states.
Because the tip points shown in fig. \ref{fig:fs}(a) have larger density of states than the valley points, 
the weights of $\vq_5$ is larger than those of $\vq_2$, resulting in the stronger features observed for $\vq_5$.
The strong features near $\vq=0$ correspond to small $\vec{q}$ scatterings around the tips and valleys points,
which have also be seen in experiments.  Our results reproduce satisfactorily the experimental findings and are also
consistent with the analysis from the spin-orientation selection rule discussed above.

As the scanning energy increases further, the surface states along the $\Gamma-M$ direction start to merge into the
conduction band of the bulk states.  In this case, the tips of the constant energy contour will be
mixed up with these bulk bands, which weakens the $\vq_5$ scattering but enhances the
small $\vq$ scatterings near the $\Gamma$ point.  This is consistent with the experiment \cite{zhang2009}, showing
that the area of the strong features near $\Gamma$ point becomes much larger after the scanning energy exceeds the bottom
of the conduction band.

\section{Spin-orbital scattering impurity}
Now we briefly comment on the effect of the spin-orbit scattering given in eq. \ref{hs} which in principle exists in any realistic system.
Since surface states of the topological insulator Bi$_2$Te$_3$ are
two-dimensional, the spin-orbit scattering potential only has one component:
\be
{\cal H}^{\rm SO}_{\rm imp}=i c V_0\!\!\int\!\! d^2k\,d^2k'\>kk'\sin(\phikp-\phik) \,\psid(\vkp)\,\sigma^z\,\psi(\vk).
\ee
Backscattering is still forbidden because of the $\sin(\phikp-\phik)$ factor.
Although $\sigma^z$ does not flip spin, the angle-dependence $\sin(\phikp-\phik)$ gives
rise to an additional suppression beyond that from the spin-orientation
selection rule discussed in the case of scalar impurity scattering.
Moreover, because the matrix element is linear in $kk'$, the spin-orbit
scattering tends to enhance the scatterings between quasiparticles
with large momenta.  All these additional effects due to the spin-orbit scattering can be
roughly seen in a straightforward calculation froim eq. \ref{vkk}:
\bea
\big|V^{(11)}_{A{\bar A}}\big|^2 &=& \big|V^{(11)}_{W{\bar W}}\big|^2 = 0  \\
\big|V^{(11)}_{AC}\big|^2 &=&\frac{V_0^2}{4}\Big[\big(1-\frac{3}{2}ck^2_L \big)^2
+3\cos^2 \vartheta \big(1+\frac{1}{2}ck^2_L\big)^2\Big]\vpha\nn\\
\big|V^{(11)}_{AB}\big|^2 &=&\frac{3V_0^2}{4}\sin^2 \vartheta\big(1-\frac{1}{2}ck^2_L\big)^2\nn\\
\big|V^{(11)}_{WX}\big|^2 &=&\frac{3V_0^2}{4}\big(1-\frac{1}{2}ck^2_U\big)^2\vpha\nn\\
\big|V^{(11)}_{WY}\big|^2 &=&\frac{V_0^2}{4}\big(1-\frac{3}{2}ck^2_U\big)^2\ . \nn 
\eea
Nonzero $c$ brings in new interferences which could lead to unusual
suppressions or enhancements for some scattering wavevectors, depending
not only on the magnitude and sign of $c$, but also on the scanning energy $E$.
In fig. \ref{fig:qpi}(b) we show the QPI image for $c=0.5$.  While the main features are still
similiar to those of fig. \ref{fig:qpi}(a), new prominent features associated with larger
momentum scatterings are visible.  Since the matrix elements for spin-orbit scattering are larger for
quasiparticles with larger momentum, this term will become more and more important as the scanning
energy $E$ increases.  A detailed analysis of the spin-orbit scattering will be presented in a future publication.
In comparison with the results in ref. \cite{zhang2009}, we find that spin-orbit scattering from the impurity of the Ag
atom is not very important in this particular experiment.

\section{Conclusion}
In conclusion, we have analyzed the quasiparticle interference
induced by nonmagnetic impurities on the surface of the topological insulator Bi$_2$Te$_3$ using
a $T$-matrix approach . While the backscattering is completely
forbidden by time-reversal symmetry, other scatterings are allowed, resulting in the QPI patterns observed in
STM experiments \cite{alpichshev2009,zhang2009} .  We have shown further that the scattering strengths depends crucially
on the spin orientations of the eigenfunctions. 
Since nonmagnetic impurities can not flip spin, the scalar scattering potential between two eigenstates is larger as their spin overlap is larger.
Combined with the variation of the density of states, we have shown that
some of the scatterings might be too weak to be seen
in comparison with the strongest ones, and our results successfully reproduce the QPI patern observed in experiments.
We have further discussed the effect of the spin-orbit scattering on the QPI pattern.
While the backscattering is still forbidden, we find that the spin-orbit scattering enhances several new features at large momentum, 
and the detailed QPI features strongly depends on the sign and strength of the spin-orbit scattering potential.

We are grateful to Xi Chen, Liang Fu, Aharon Kapitulnik, Qin Liu, Xiaoliang Qi, Qikun Xue for insightful discussions. CW and WCL
are supported by ARO-W911NF0810291. S CZ is supported by the
Department of Energy, Office of Basic Energy Sciences, Division of
Materials Sciences and Engineering, under contract
DE-AC02-76SF00515.

{\it Note added} -- While this paper was about completion, we learned a related work by Zhang {\it et al.}\cite{zhang20092}.


\end{document}